\documentstyle[11pt]{article}
\begin{document}
\hfill{UTTG-12-05}

\vspace{12pt}

\begin{center}
{\bf Living in the Multiverse}\\

\vspace{12pt}

Opening Talk at the Symposium "Expectations of a Final Theory" at Trinity College, Cambridge, September 2, 2005; to be published in {\it Universe or Multiverse?}, ed. B. Carr (Cambridge University Press).\\

\vspace{12pt}

Steven Weinberg\\
Physics Department, University of Texas at Austin
\end{center}

\vspace{12pt}

Most advances in the history of science have been marked by discoveries about nature, but at certain turning points we have made discoveries about science itself.  These discoveries lead to changes in how we score our work, in what we consider to be an acceptable theory.

For an example look back to a discovery  made just one hundred years ago.  As you recall, before 1905 there had been numerous unsuccessful efforts to detect changes in the speed of light due to the motion of the earth through the ether.  Attempts were made by Fitzgerald, Lorentz, and others to construct a mathematical model of the electron (which was then conceived to be the chief constituent of all matter),  that would explain how rulers contract when moving through the ether in just the right way to keep the apparent speed of light unchanged.  Einstein instead offered a symmetry principle, which stated that not just the speed of light but all the laws of nature are unaffected by a transformation to a frame of reference in uniform motion.  Lorentz grumbled that Einstein was simply assuming what he and others had been trying to prove.  But history was on Einstein's side.  The 1905 Special Theory of Relativity was the beginning of a general acceptance of symmetry principles as a valid basis for physical theories.  

This was how Special Relativity made a change in science itself.  From one point of view, Special Relativity was no big thing --- it just amounted to the replacement of one 10 parameter spacetime symmetry group, the Galileo group, with another 10 parameter group, the Lorentz group.  But never before had a symmetry principle been taken as a legitimate hypothesis on which to base a physical theory.

As usually happens with this sort of revolution, Einstein's advance came with a retreat in another direction: The effort to construct a model of the electron was suspended for decades. Instead, symmetry principles increasingly became the dominant foundation for physical theories.  This tendency was accelerated after the advent of quantum mechanics in the 1920s, because the survival of symmetry principles in quantum theories imposes highly restrictive consistency conditions (existence of antiparticles, connection between spin and statistics, cancelation of infinities and anomalies) on physically acceptable theories.  Our present Standard Model of elementary particle interactions can be regarded as simply the consequence of certain gauge symmetries and the associated quantum mechanical consistency conditions.  

The development of the Standard Model did not involve any changes in our conception of what was acceptable as a basis for physical theories.  Indeed, the Standard Model can be regarded as just quantum electrodynamics writ large.  Similarly, when the effort to extend the Standard Model to include gravity led to widespread interest in string theory, we expected to score the success or failure of this theory in the same way as for the Standard Model: String theory would be a success if its symmetry principles and consistency conditions led to a successful prediction of the free parameters of the Standard Model.

Now we may be at a new turning point, a radical change in what we accept as a legitimate foundation for a physical theory.  The current excitement is is of course a consequence of the discovery of a vast number of solutions of string theory, beginning in 2000 with the work of Bousso and Polchinski.\footnote{R. Bousso and J. Polchinski, JHEP {\bf 0006}, 006 (2000).}  The compactified six dimensions in Type II string theories typically have a large number (tens or hundreds) of topological fixtures (3-cycles), each of which can be threaded by a variety of fluxes.  The logarithm of the number of allowed sets of values of these fluxes is proportional to the number of topological fixtures.  Further, for each set of fluxes one obtains a different effective field theory for the modular parameters that describe the compactified 6-manifold, and for each effective field theory the number of local minima of the potential for these parameters is again proportional to the number of topological fixtures.  Each local minimum corresponds to the vacuum of a possible stable or metastable universe.

Subsequent work by  Giddings, Kachru, Kallosh, Linde, Maloney, Polchinski, Silverstein, Strominger, and Trivedi (in various combinations\footnote{S. B. Giddings, S. Kachru, and J. Polchinski, Phys. Rev. {\bf D66}, 106006 (2002); A. Maloney, E. Silverstein, and A. Strominger, hep-th/0205316; S. Kachru, R. Kallosh, A. D. Linde, and S. P. Trivedi, Phys. Rev. {\bf D68}, 046005 (2003).})
established the existence of a large number of vacua with positive energy densities.  Ashok and Douglas\footnote{S. K. Ashok and M. Douglas, JHEP {\bf 0401}, 060 (2004).} estimated  the number of these vacua to be of order $10^{100}$ to  $10^{500}$.  Susskind\footnote{L. Susskind, hep-th/0302219} gave the name ``string landscape'' to this multiplicity of vacua, taking the term from biochemistry, where the possible choices of orientation of each chemical bond in large molecules leads to a vast number of possible configurations.
Unless one can find a reason to reject all but a few of the string theory vacua, we will have to accept that much of what we had hoped to calculate are environmental parameters, like the distance of the earth from the sun, whose values we will never be able to deduce from first principles.

We lose some, and win some.  The larger the number of possible values of physical parameters provided by the string landscape, the more string theory legitimates anthropic reasoning as a new basis for physical theories: Any scientists who study nature must live in a part of the landscape where physical parameters take values suitable for the appearance of life and its evolution into scientists.  

An apparently successful example of anthropic reasoning was already at hand by the time the string landscape was discovered.  For decades there seemed to be something peculiar about the value of the vacuum energy $\rho_V$.  Quantum fluctuations in known fields at well-understood energies (say, less than 100 GeV) give a value of $\rho_V$ larger than observationally allowed by a factor $10^{56}$.  This contribution to the vacuum energy might be canceled by quantum fluctuations of higher energy, or by simply including a suitable cosmological constant term in the Einstein field equations, but the cancelation would have to be exact to 56 decimal places.  No symmetry argument or adjustment mechanism could be found that would explain such a cancelation.  Even if such an explanation could be found, there would be no reason to suppose that the remaining net vacuum energy would be comparable to the {\em present} value of the matter density, and since it is certainly not very much larger, it was natural to suppose that it is very much less, too small to be detected.

On the other hand, if $\rho_V$ takes a broad range of values in the multiverse, then it is natural for scientists to find themselves in a subuniverse in which $\rho_V$ takes a value suitable for the appearance of scientists.  I pointed out in 1987 that this value for $\rho_V$ 
can't be too large and positive, because then galaxies and stars would not form.\footnote{S. Weinberg, Phys. Rev. Lett. {\bf 59}, 2607 (1987).}  Roughly, this limit is that $\rho_V$ should be less than the mass density of the universe at the time when galaxies first condense.  Since this was in the past, when the mass density was larger than at present, the anthropic upper limit on the vacuum energy density  is larger than the present mass density, but not many orders of magnitude greater.

But anthropic arguments provide not just a bound on $\rho_V$; they give us some idea of the value to be expected: $\rho_V$ should be not very different from the mean of the values suitable for life.  This is what  Vilenkin\footnote{A. Vilenkin, Phys. Rev. Lett. {\bf 74}, 846 (1995)} calls the ``principle of mediocrity.'' This mean is positive, because if $\rho_V$ were negative  it would have to be less in absolute value than the mass density of the universe during the whole time that life evolves, since otherwise the universe would collapse before any astronomers come on the scene,\footnote{J. D. Barrow and F. J. Tipler, {\em The Anthropic Cosmological Principle} (Clarendon, Oxford, 1986).} while if positive $\rho_V$ only has to be less than the mass density of the universe at the time when most galaxies form, giving a much broader range of possible positive than negative values.  In 1997-8 Martel, Shapiro, and I\footnote{H. Martel, P. Shapiro, and S. Weinberg, Astrophys. J. {\bf 492}, 29 (1998).  For earlier calculations,  see G. Efstathiou, Mon. Not. Roy. Astron. Soc. {\bf 274}, L73 (1995); S. Weinberg, in {\em Critical Dialogues in Cosmology}, ed. N. Turok (World Scientific, 1997).}  carried out a detailed calculation of the probability distribution of values of $\rho_V$ seen by astronomers throughout the multiverse, under the assumption that the {\em a priori} probability distribution is flat in the relatively very narrow range that is anthropically allowed.  At that time the value of the primordial rms fractional density fluctuation $\sigma$ was not well known, since the value inferred from observations of the cosmic microwave background depended on what one assumed for $\rho_V$.  It was therefore not possible to calculate a mean expected value of $\rho_V$, but for any assumed value of $\rho_V$ we could estimate $\sigma$ and use the result to calculate the fraction of astronomers that would observe a value of $\rho_V$ as small as the assumed value.  In this way we concluded that if $\Omega_\Lambda$ turned out to be much less than 0.6, anthropic reasoning could not explain why it was so small.  The editor of the Astrophysical Journal objected to publishing papers about anthropic calculations, and we had to sell our article by pointing out that we had provided a strong argument for abandoning an anthropic explanation of a small value of $\rho_V$, if it turned out to be too small.

Of course, it turned out that $\rho_V$ is not too small.  Soon after this work, observations of type Ia supernovae revealed that the expansion of the universe is accelerating,\footnote{A. G. Riess {\em et al.}, Astron. J. {\bf 116}, 1009 (1998); S. Perlmutter {\em et al.}, Astrophys. J. {\bf 517}, 565 (1999).} and gave the result that $\Omega_V\simeq 0.7$.  In other words the ratio of the vacuum energy density to the present mass density $\rho_{M0}$ in {\em our} subuniverse (which I use just as a convenient measure of density)  is about 2.3, a conclusion subsequently confirmed by observations of the microwave background.\footnote{WMAP collaboration, Astrophys. J. Suppl. {\bf 148} (2003).}  

This is still a bit low.  Martel, Shapiro, and I had found that the probability of a vacuum energy density this small was 12\%.  I have now recalculated the probability distribution, using WMAP data and a better transfer function, with the result that the probability of a random astronomer seeing a value as small as $2.3\rho_{M0}$ is increased to 15.6\%.  Now that we know $\sigma$, we can also calculate that the  median vacuum energy density is $13.3\rho_{M0}$.

I should mention  a complication in these calculations.  The average of the product of density fluctuations at different points becomes infinite as these points approach each other, so the rms fractional density fluctuation $\sigma$ is actually infinite.  Fortunately, it is not $\sigma$ itself that is really needed in these calculations, but the rms fractional density fluctuation averaged over a sphere of co-moving radius $R$ taken large enough so that the density fluctuation is able to hold on efficiently to the heavy elements produced in the first generation of stars.  The results mentioned above were calculated for $R$ (projected to the present) equal to 2 Mpc.  These results are rather sensitive to the value of $R$; for $R=1$ Mpc, the probability of finding a vacuum energy as small as $2.3\rho_{M0}$ is only 7.2\%.  The estimate of the required value of $R$ involves complicated astrophysics, and needs to be better understood.

Now I want to take up four problems we have to face in working out the anthropic implications of the string landscape.

\vspace{12pt}

\noindent
{\bf I What is the shape of the string landscape?}

Douglas\footnote{M. R. Douglas, hep-ph/0401004; Compt. Rend. Phys. {\bf 5}, 965 (2004).} and Dine\footnote{M. Dine, D. O'Neil and Z. Sun, JHEP {\bf 0507}, 014 (2005); M. Dine and Z. Sun, hep-th/0506246.} and their co-workers have taken the first steps in finding the statistical rules governing different string vacua.  I can't comment usefully on this, except to say that it wouldn't hurt in this work if we knew what string theory is.

\vspace{12pt}

\noindent
{\bf II What constants scan?}

Anthropic reasoning makes sense for a given constant if  the  range over which the constant varies in the landscape is large compared with the anthropically allowed range of values of the constant, for then it is reasonable to assume that the {\em a priori} probability distribution is flat in the anthropically allowed range.   We need to know what constants  actually ``scan'' in this sense.   Physicists would like to be able to calculate as much as possible, so we hope that not too many constants scan.

The most optimistic hypothesis is that the only constants that scan are the few whose dimensionality is a positive power of mass: the vacuum energy, and whatever scalar mass or masses set the scale of electroweak symmetry breaking.  With all other parameters of the Standard Model fixed, the scale of electroweak symmetry breaking is bounded by about 1.4 to 2.7 times its value in our subuniverse, by the condition that the pion mass should be small enough to make the nuclear force strong enough to keep the deuteron stable against fission.\footnote{V.Agrawal, S. M. Barr, J. F. Donoghue, and D. Seckel, Phys. Rev. {\bf D 57}, 5480 (1998).}  (The condition that the deuteron be stable against beta decay, which yields a tighter bound, does not seem to me to be necessary.  Even a beta-unstable deuteron would live long enough to allow cosmological helium synthesis; helium would be burned to heavy elements in the first generation of very massive stars; and then subsequent generations could have long lifetimes burning hydrogen through the carbon cycle.)   But the mere fact that the electroweak symmetry breaking scale is only a few orders of magnitude larger than the QCD scale should not in itself lead us to conclude that it must be anthropically fixed.  There is always the possibility that the electroweak symmetry breaking scale is determined by the energy at which some gauge coupling constant becomes strong, and if that coupling happens to grow with decreasing energy a little faster than the QCD coupling then the electroweak breaking scale will naturally be a few orders of magnitude larger than the QCD scale.

If the electroweak symmetry breaking scale is anthropically fixed, then we can give up the decades long search for a natural solution of the hierarchy problem.  This is a very attractive prospect, because none of the ``natural'' solutions that have been proposed, such as technicolor or low energy supersymmetry, were ever free of difficulties.  In particular, giving up low energy supersymmetry can restore some of the most attractive features of the non-supersymmetric standard model: automatic conservation of baryon and lepton number in interactions up to dimension 5 and 4, respectively; natural conservation of flavors in neutral currents; and a small neutron electric dipole moment.  Arkani-Hamed and Dimopoulos\footnote{N. Arkani-Hamed and S. Dimopoulos, JHEP {\bf 0506}, 073 (2005).  Also see G. F. Giudice and A. Romanino, Nucl. Phys. B {\bf 699}, 65 (2004); N. Arkani-Hamed, S. Dimopoulos, G. F. Giudice, and A. Romanino, Nucl. Phys. B {\bf 709}, 3 (2005); A. Delgado and G. F. Giudice, hep-ph/0506217.} have even shown how it is possible to keep the good features of supersymmetry, such as a more accurate convergence of the $SU(3)\times SU(2)\times U(1)$ couplings to a single value, and the presence of candidates for dark matter WIMPs.  The idea of this ``split supersymmetry'' is that, although supersymmetry is broken at some very high energy, the gauginos and higgsinos are kept light by a chiral symmetry.  [An additional discrete symmetry is needed to prevent lepton-number violation in higgsino-lepton mixing, and to keep the lightest supersymmetric particle stable.]  One of the nice things about split supersymmetry is that, unlike many of the things we talk about these days, it makes predictions that can be checked when the LHC starts operation.  One expects a single neutral Higgs with a mass in the range 120 to 165 GeV, possible winos and binos but no squarks or sleptons, and a long-lived gluino.  (Incidentally, a Stanford group\footnote{A. Arvanitaki, C. Davis, P. W. Graham, A. Pierce, and J. G. Wacker, hep-ph/0504210.} has recently used considerations of big bang nucleosynthesis to argue that a 1 TeV gluino must have  a lifetime less than 100 seconds, indicating a supersymmetry breaking scale less than $10^{10}$ GeV, which might create problems for proton stability.  But I wonder whether, even if the gluino has a longer lifetime and  decays after nucleosynthesis, the universe might not thereby be reheated above the temperature of helium dissociation, giving big bang nucleosynthesis a second chance to produce the observed helium abundance.)

What about the dimensionless Yukawa couplings of the Standard Model?
Hogan\footnote{C. Hogan, Rev. Mod. Phys. {\bf 72}, 1149 (2000); and astro-ph/0407086.}   has analyzed the anthropic constraints on these couplings, with the electroweak symmetry breaking scale and the sum of the $u$ and $d$ Yukawa couplings held fixed, to avoid complications due to the dependence of nuclear forces on the pion mass.  He imposes the conditions that (1) $m_d-m_u-m_e>1.2$ MeV, so that the early universe doesn't become all neutrons; (2) $m_d-m_u+m_e<3.4$ MeV, so that the $pp$ reaction is exothermic, and (3) $m_e>0$.  With three conditions on the two parameters $m_u-m_d$ and $m_e$, he naturally finds these parameters are limited to a finite region, which turns out to be quite small.  At first sight, this gives the impression that the quark and lepton Yukawa couplings are subject to stringent anthropic constraints, in which case we might infer that the Yukawa couplings probably scan.  

I have two reservations about this conclusion.  The first reservation is that the $pp$ reaction is not necessary for life.  For one thing, the 
$pep$ reaction $p+p+e^-\rightarrow d+\nu$ can keep stars burning hydrogen for a long time.  For this, we do not need $m_d-m_u+m_e<3.4$ MeV, but only the weaker condition $m_d-m_u-m_e<3.4$ MeV.  The three conditions then do not constrain $m_d-m_u$ and $m_e$ separately to any finite region, but only constrain the single parameter $m_d-m_u-m_e$ to lie between 1.2 MeV and 3.4 MeV, not a very tight anthropic constraint.
(In fact, He$^4$ will be stable as long as $m_d-m_u-m_e$ is less than about 13 MeV, so stellar nucleosynthesis can begin with helium burning in the heavy stars of Population III, followed by hydrogen burning in later generations of stars.)  My second reservation is that the anthropic constraints on the Yukawa couplings are alleviated if we suppose (as discussed above) that the electroweak symmetry breaking scale is not fixed, but free to take whatever value is anthropically necessary.  For instance, according to the results of reference 13, the deuteron binding energy could be made as large as about 3.5 MeV by taking the electroweak breaking scale much less than it is in our universe, in which case even the condition that the $pp$ reaction be exothermic becomes much looser.

Incidentally, I don't set much store by the famous ``coincidence'' emphasized by Hoyle, that there is an excited state of C$^{12}$ with just the right energy to allow carbon production via $\alpha$--Be$^8$ reactions in stars.  We know that even--even nuclei have states that are well described as composites of $\alpha$-particles.  One such state is the ground state of Be$^8$, which is unstable against fission into two alpha particles.  The same $\alpha$-$\alpha$ potential that produces that sort of unstable state in Be$^8$ could naturally be expected to produce an unstable state in C$^{12}$ that is essentially a composite of three $\alpha$ particles, and that therefore appears as a low-energy resonance in $\alpha$--Be$^8$ reactions.  So the existence of this state doesn't seem to me to provide any evidence of fine tuning.

What else scans?  Tegmark and Rees\footnote{M. Tegmark and M. J. Rees, Astrophys. J. {\bf 499}, 526 (1998).} have raised the question whether the rms density fluctuation $\sigma$ may itself scan.  If it does, then the anthropic constraint on the vacuum energy becomes weaker, resuscitating to some extent the problem of why $\rho_V$ is so small.  But Garriga and Vilenkin\footnote{J. Garriga and A. Vilenkin, hep-th/0508005.} have pointed out that it is really $\rho_V/\sigma^3$ that is constrained anthropically, so that even if $\sigma$ does scan the anthropic prediction of this ratio remains robust.

Arkani-Hamed, Dimopoulos, and Kachru\footnote{N. Arkani-Hamed, S. Dimopoulos, and S. Kachru, hep-th/0501082, referred to below as ADK.} have offered a possible reason to suppose that most constants do not scan.  If there are a large number $N$ of decoupled modular fields, each taking a few possible values, then the probability distribution of quantities that depend on all these fields will be sharply peaked, with a width proportional to $1/\sqrt{N}$.  According to Distler and Varadarajan,\footnote{J. Distler and U. Varadarajan, hep-th/0507090.} it is not really necessary here to make arbitrary assumptions about the decoupling of the various scalar fields; it is enough to adopt the most general polynomial superpotential that is stable, in the sense that radiative corrections do not change the effective couplings for large $N$ by amounts larger than the couplings themselves.  Distler and Varadarajan emphasize cubic superpotentials, because polynomial superpotentials of order higher than cubic presumably make no physical sense.  But it is not clear that even cubic superpotentials can be plausible approximations, or that peaks will occur at reasonable values in the distribution of dimensionless couplings rather than of some combinations of these couplings.\footnote{M. Douglas, private communication.}  It also is not clear that the multiplicity of vacua in this kind of effective scalar field theory can properly represent the multiplicity of flux values in string theories,\footnote{T. Banks, hep-th/0011255.} but even if not, it presumably can represent the variety of minima of the potential for a given set of flux vacua.

If most constants do not effectively scan, then why should anthropic arguments work for the vacuum energy and the electroweak breaking scale? ADK point out that, even if some constant has a relatively narrow distribution, anthropic arguments will still apply if the anthropically allowed range is even narrower and near a point around which the distribution is symmetric.  (ADK suppose that this point would be at zero, but this is not necessary.)  This is the case, for instance, for the vacuum energy if the superpotential $W$ is the sum of the superpotentials $W_n$ for a large number of decoupled scalar fields, for each of which there is a separate broken $R$ symmetry, so that the possible values of each $W_n$ are equal and opposite.  The probability distribution of the total superpotential $W=\sum_{n=1}^N W_n$ will then be a Gaussian peaked at $W=0$ with a width proportional to $1/\sqrt{N}$, and the probability distribution of the supersymmetric vacuum energy $-8\pi G|W|^2$ will extend over a correspondingly narrow range of negative values, with a maximum at zero.  When supersymmetry breaking is taken into account, the probability distribution widens to include positive values of the vacuum energy, extending out to a positive value depending on the scale of supersymmetry breaking.  For any reasonable supersymmetry breaking scale, this probability distribution, though narrow compared with the Planck scale, will be very wide compared with the vary narrow anthropically allowed range around $\rho_V=0$, so within this range the probability distribution can be expected to be flat, and anthropic arguments should work.  Similar remarks apply to the $\mu$-term of the supersymmetric Standard Model, which sets the scale of electroweak symmetry breaking.

\vspace{12pt}

\noindent
{\bf III. How should we calculate anthropically conditioned probabilities?}

We would expect the anthropically conditioned probability distribution for a given value of any constant that scans to be proportional to the number of scientific civilizations that observe that value.  In the calculations described above, Martel, Shapiro, and I took this number to be proportional to the {\em fraction} of baryons that find themselves in galaxies, but what if the total number of baryons itself scans?  What if it is infinite?  

\vspace{12pt}

\noindent
{\bf IV. How is the landscape populated?}

There are at least four ways in which we might imagine the different ``universes'' described by the string landscape actually to exist:

\begin{enumerate}
\item The  various subuniverses may be simply different regions of space.  This is most simply realized in the chaotic inflation theory.\footnote{A. D. Linde, Phys. Lett. {\bf 129B}, 177 (1983); A. Vilenkin, Phys. Rev.  {\bf D 27}, 2848 (1983); A. D. Linde, Phys. Lett. B {\bf 175}, 305 (1986); Phys. Scripta {\bf T15}, 100 (1987); Phys. Lett. B{\bf 202}, 194 (1988).}  The scalar fields in different inflating patches may take different values, giving rise to different values for various effective coupling constants.    Indeed, Linde speculated about the application of the anthropic principle to cosmology soon after the proposal of chaotic inflation.\footnote{A. D. Linde, in {\it The Very Early Universe}, ed. G. W. Gibbons, S. W. Hawking, and S. Siklos (Cambridge University Press, 1983); Rept. Progr. Phys. {\bf 47}, 925 (1984).}

\item The subuniverses may be different eras of time  in a single big bang.  For instance, what appear to be constants of nature might actually depend on scalar fields that change very slowly as the universe expands.\footnote{T. Banks, Nucl. Phys. B {\bf 249}, 332 (1985).}  
\item The subuniverses may be different regions of spacetime.  This can happen if, instead of changing smoothly with time, various scalar fields on which the ``constants'' of nature depend  change in a sequence of first-order phase transitions.\footnote{L. Abbott, Phys. Lett. {\bf B150}, 427 (1985); J. D. Brown and C. Teitelboim, Phys. Lett. B {\bf 195}, 177 (1987); Nucl. Phys. B {\bf 297}, 787 (1987).}  In these transitions  metastable bubbles  form within a region of higher vacuum energy; then within each bubble there form further  bubbles of even lower vacuum energy; and so on.  In recent years this idea has been revived in the context of the string landscape.\footnote{R. Bousso and J. Polchinski, {\em op. cit.}; J. L. Feng, J. March-Russel, S. Sethi, and F. Wilczek, Nucl. Phys. B {\bf 602}, 307 (2001); H. Firouzjahi, S. Sarangji, and S.-H. Henry Tye, JHEP {\bf 0409}, 060 (2004); B.Freivogel, M. Kleban, M. R. Martinez, and L. Susskind, hep-th/0505232.}
\item The subuniverses could be different parts of quantum mechanical Hilbert space.  In a reinterpretation of Hawking's earlier work on the wave function of the universe,\footnote{S. W. Hawking, Nucl. Phys. B {\bf 239},  257 (1984); and in {\em Relativity, Groups, and Topology II}, NATO Advanced Study Institute Session XL, Les Houches, 1983, ed. B.S. DeWitt and R. Stora (Elsevier, Amsterdam, 1984): p. 336.  Some of this work is based on an initial condition for the origin of the universe proposed by J. Hartle and S. W. Hawking, Phys. Rev. D {\bf 28}, 2960 (1983).}  Coleman\footnote{S. Coleman, Nucl. Phys. B {\bf 307}, 867 (1988). It has been argued  that the wave function of the universe is sharply peaked at values of the constants that yield a zero vacuum energy at late times, by S. W. Hawking, in {\em Shelter Island II --- Proceedings of the 1983 Shelter Island Conference on Quantum Field Theory and the Fundamental Problems of Physics}, ed. R. Jackiw {\em et al.} (MIT  Press, Cambridge, 1985);  Phys. Lett. B {\bf 134}, 403 (1984); E. Baum, Phys. Lett. B {\bf 133}, 185 (1984); S. Coleman, Nucl. Phys. B {\bf 310}, 643 (1985). This view has been challenged; see W. Fischler, I. Klebanov, J. Polchinski, and L. Susskind, Nucl. Phys. B {\bf 237}, 157 (1989).  I am assuming here that there are no such peaks.} showed that certain topological fixtures known as wormholes in the path integral for the Euclidean wave function of the universe would lead to a superposition of wave functions in which any coupling constant not constrained by symmetry principles would take any possible value.  Ooguri, Vafa, and Verlinde\footnote{H. Ooguri, C. Vafa, and E. Verlinde, hep-th/0502211.} have argued for a particular wave function of the universe, but it escapes me how anyone can tell whether this or any other proposed wave function is {\em the} wave function of the universe.

\end{enumerate}

These alternatives are by no means mutually exclusive.  In particular, it seems to me that, whatever one concludes about alternatives 1, 2, and 3, we will still have the possibility that the wave function of the universe is a superposition of different terms representing different ways of populating the landscape in space and/or time.

\vspace{12pt}

In closing,  I would like to comment about the impact of anthropic reasoning within and beyond the physics community.  Some physicists have expressed a strong distaste for anthropic arguments.  (I have heard David Gross say  ``I hate it.'')  This is understandable.  Theories based on anthropic calculation certainly represent a retreat from what we had hoped for: the calculation of all fundamental parameters from first principles.  It is too soon to give up on this hope, but without loving it we may just have to resign ourselves to a retreat, just as Newton had to give up Kepler's hope of a calculation of the relative sizes of planetary orbits from first principles.  

There is also a less creditable reason for hostility to the idea of a multiverse, based on the fact that we will never be able to observe any subuniverses except our own.  Livio and Rees\footnote{M. Livio and M. J. Rees, Science {\bf 309}, 1022 (12 August, 2003).} and Tegmark\footnote{M. Tegmark, Ann. Phys. {\bf 270}, 1 (1998).} have given  thorough discussions of various other ingredients of accepted theories that we will never be able to observe, without our being led to reject these theories.  The test of a physical theory is not that everything in it should be observable and every prediction it makes should be testable, but rather that enough is observable and enough predictions are testable to give us confidence that the theory is right.

Finally, I have heard the objection that, in trying to explain why the laws of nature are so well suited for the appearance and evolution of life, anthropic arguments take on some of the flavor of religion.  I think that just the opposite is the case.  Just as Darwin and Wallace  explained how the wonderful adaptations of living forms could arise without supernatural intervention, so the string landscape may explain how the constants of nature that we observe can take values suitable for life without being fine-tuned by a benevolent creator.  I found this parallel well understood in a surprising place, a New York Times op-ed article by Christoph Sch\"{o}nborn, Cardinal Archbishop of Vienna.\footnote{C. Sch\"{o}nborn, N. Y. Times, 7 July 2005, p. A23.}  His article concludes as follows:
\begin{quotation}
``Now, at the beginning of the 21st century, faced with scientific claims like neo-Darwinism and the multiverse hypothesis in cosmology invented to avoid the overwhelming evidence for purpose and design found in modern science, the Catholic Church will again defend human nature by proclaiming that the immanent design evident in nature is real.  Scientific theories that try to explain away the appearance of design as the result of `chance and necessity' are not scientific at all, but, as John Paul put it, an abdication of human intelligence.''
\end{quotation}
It's nice to see work in cosmology get some of the attention given these days to evolution, but of course it is not religious preconceptions like these that can decide any issues in science.

It must be acknowledged that there is a big difference in the degree of confidence we can have in neo-Darwinism and in the multiverse.  It is settled, as well as anything in science is ever settled, that the adaptations of living things on earth have come into being through natural selection acting on random undirected inheritable variations.  About the multiverse, it is appropriate to keep an open mind, and opinions among scientists differ widely.  In the Austin airport on the way to this meeting I noticed for sale the October issue of a magazine called {\em Astronomy}, having on the cover the headline ``Why You Live in Multiple Universes.''  Inside I found a report of a discussion at a conference at Stanford, at which Martin Rees said that he was sufficiently confident about the multiverse to bet his dog's life on it, while Andrei Linde said he would bet his own life.  As for me, I have just enough confidence about the multiverse to bet the lives of both Andrei Linde {\em and} Martin Rees's dog.

\vspace{12pt}`

This material is based upon work supported by the National Science Foundation under Grants Nos. PHY-0071512 and PHY-0455649 and with support from The Robert A. Welch Foundation, Grant No. F-0014, and also grant support from the US Navy, Office of Naval Research, Grant Nos. N00014-03-1-0639 and N00014-04-1-0336, Quantum Optics Initiative.

\end{document}